\documentclass[journal]{IEEEtran}
\usepackage{bibentry}
\usepackage{xcolor,soul,framed} %,caption

\colorlet{shadecolor}{yellow}
\usepackage[pdftex]{graphicx}
\graphicspath{{../pdf/}{../jpeg/}}
\DeclareGraphicsExtensions{.pdf,.jpeg,.png}

\usepackage[cmex10]{amsmath}
\usepackage{array}
\usepackage{mdwmath}
\usepackage{mdwtab}
\usepackage{eqparbox}
\usepackage{url}
\usepackage{multirow}

\usepackage{algorithmic}
\usepackage{algorithm}

\begin{document}
\bstctlcite{IEEEexample:BSTcontrol}
\title{Exploiting Shared Knowledge from Non-COVID Lesions for Annotation-Efficient COVID-19 CT Lung Infection Segmentation}
\author{Yichi Zhang, Qingcheng Liao, Lin Yuan, He Zhu, Jiezhen Xing, and Jicong Zhang
	\thanks{This work is supported by the National Key Research and Development Program of China (2016YFF0201002), the University Synergy Innovation Program of Anhui Province (GXXT-2019-044), and the National Natural Science Foundation of China (61301005). The first two arthors contributed equally to this work.}
	\thanks{Yichi Zhang, Qingcheng Liao and Jiezhen Xing are with School of Biological Science and Medical Engineering, Beihang University, Beijing, China.}
	\thanks{Lin Yuan is with College of Biomedical Engineering, Taiyuan University of Technology, Taiyuan, China.}
	\thanks{He Zhu is with School of Computer Science and Engineering, Beihang University, Beijing, China.}
	\thanks{Jicong Zhang (corresponding author) is with School of Biological Science and Medical Engineering, Beihang University, Beijing, China, and with Hefei Innovation Research Institute, Beihang University, Hefei, China, and with Beijing Advanced Innovation Centre for Biomedical Engineering, Beijing, China, and with Beijing Advanced Innovation Centre for Big Data-Based Precision Medicine, Beijing, China. (e-mail: jicongzhang@buaa.edu.cn)}
}

% ====================================================================
\maketitle

% === ABSTRACT ====================================================================
% =================================================================================
\begin{abstract}
The novel Coronavirus disease (COVID-19) is a highly contagious virus and has spread all over the world, posing an extremely serious threat to all countries. Automatic lung infection segmentation from computed tomography (CT) plays an important role in the quantitative analysis of COVID-19. 
However, the major challenge lies in the inadequacy of annotated COVID-19 datasets. Currently, there are several public non-COVID lung lesion segmentation datasets, providing the potential for generalizing useful information to the related COVID-19 segmentation task. 
In this paper, we propose a novel relation-driven collaborative learning model to exploit shared knowledge from non-COVID lesions for annotation-efficient COVID-19 CT lung infection segmentation. 
The model consists of a general encoder to capture general lung lesion features based on multiple non-COVID lesions, and a target encoder to focus on task-specific features based on COVID-19 infections. Features extracted from the two parallel encoders are concatenated for the subsequent decoder part.
We develop a collaborative learning scheme to regularize feature-level relation consistency of given input and encourage the model to learn more general and discriminative representation of COVID-19 infections. 
Extensive experiments demonstrate that trained with limited COVID-19 data, exploiting shared knowledge from non-COVID lesions can further improve state-of-the-art performance with up to 3.0\% in dice similarity coefficient and 4.2\% in normalized surface dice.
In addition, experimental results on large scale 2D dataset with CT slices show that our method significantly outperforms cutting-edge segmentation methods on all evaluation metrics.
Our proposed method promotes new insights into annotation-efficient deep learning for COVID-19 infection segmentation and illustrates strong potential for real-world applications in the global fight against COVID-19 in the absence of sufficient high-quality annotations.

\end{abstract}

% === KEYWORDS ====================================================================
% =================================================================================
\begin{IEEEkeywords}
COVID-19, Computed Tomography, Lung Infection Segmentation, Few-shot Learning, Knowledge Transfer.\\ \\
\end{IEEEkeywords}

\IEEEpeerreviewmaketitle

\section{Introduction}
\IEEEPARstart{S}{ince} the beginning of 2020, the novel coronavirus disease (COVID-19) has rapidly spread worldwide, posing an extremely serious threat and challenge to all countries. This severe disease has been declared as a public health emergency of international concern by the World Health Organization (WHO), which has caused more than 2,600,000 deaths until the date of $9_{th}$ March 2021, according to the statistics of Johns Hopkins Coronavirus Resource Center\footnote{https://coronavirus.jhu.edu/map.html}. 

As one of the most commonly used imaging methods, computed tomography (CT) plays an important role in the fight against COVID-19 \cite{mazzone2020management,oudkerk2020diagnosis,li2020coronavirus}. Researchers have proved that CT images have strong ability to capture typical features like ground glass and bilateral patchy shadows of affected patients \cite{chung2020ct} and are shown to be more sensitive compared with standard viral nucleic acid detection using real-time polymerase chain reaction (RT-PCR) for the early diagnosis of COIVD-19 infection \cite{fang2020sensitivity}. Besides, CT images can provide visual evaluation of the extent of lung abnormalities and assist the process of prognostic \cite{li2020ct}.

\begin{figure}[t]
	\includegraphics[width=9cm]{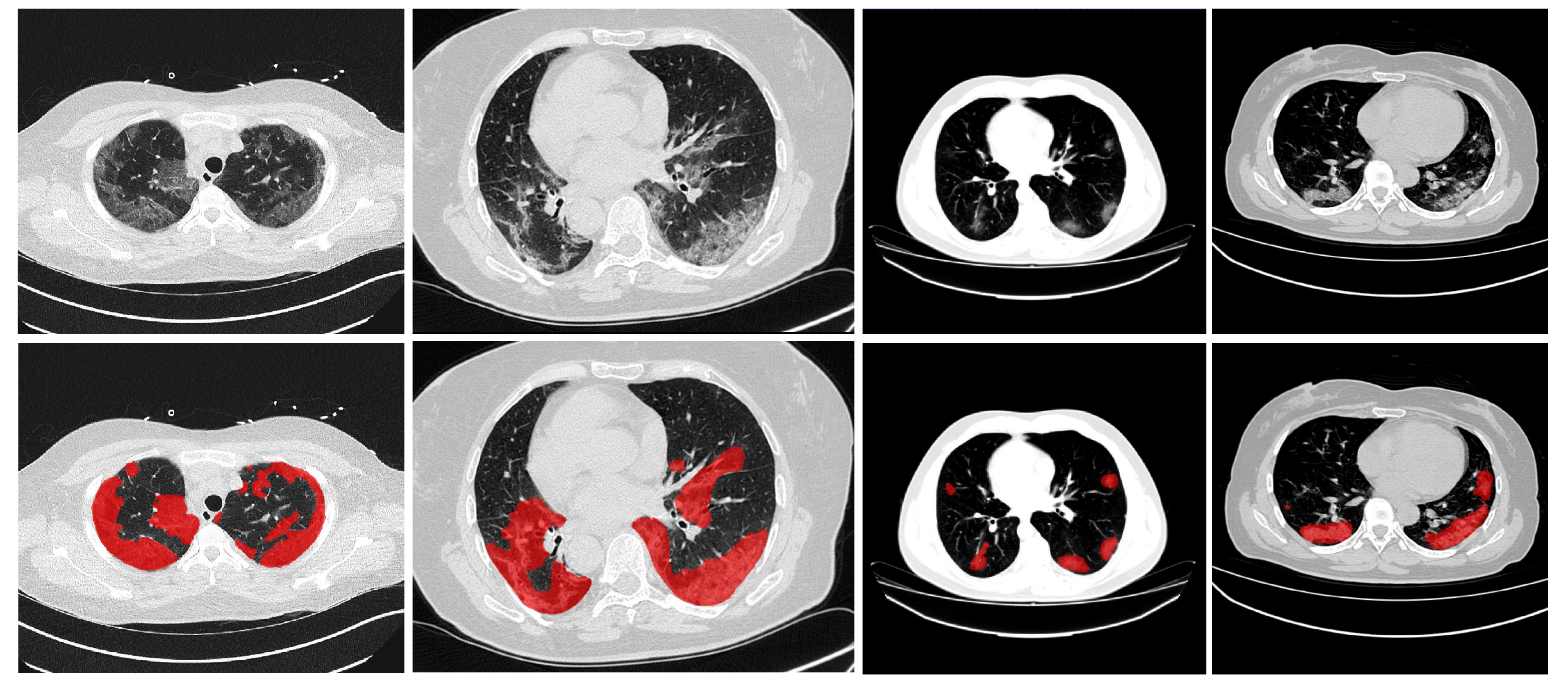}
	\centering
	\caption{Examples of COVID-19 infections in CT volumes showing the large  variations of shape, size and position of lung infections. The upper row shows the raw images (from axial view) and the lower row shows corresponding annotations of infection areas.}
	\label{covid}
\end{figure}

In clinical practice, the segmentation of lung infections from CT images is an important component to assist in further assessment and quantification of the diseases \cite{shi2020review}. Since manual contour delineation is time-consuming and laborious, and suffers from inter and intra-observer variabilities \cite{joskowicz2019inter}, it is of great significance to develop artificial intelligence-based approaches to assist in the automatic segmentation of COVID-19 infections. Recently, the unprecedented development in deep learning has showed significant improvements and achieved state-of-the-art performances in many medical image segmentation tasks \cite{bilic2019liver,heller2019state,bernard2018deep,ma2021abdomenct}, and deep neural networks have been widely applied in the global fight against COVID-19 \cite{huang2020serial, sun2020adaptive, zhang2020clinically, wang2020contrastive}.

However, the success of deep learning methods mainly requires large amount of high-quality annotated datasets, while it is impractical to collect large amount of well annotated data in real clinical approach, especially when radiologists are busy fighting the coronavirus disease. Additionally, as shown in Fig.\ref{covid}, the large variations in shape, size and position of lung infections and large inter-case variations pose great challenges for the segmentation tasks \cite{ouyang2020dual}. Therefore, exploring annotation-efficient COVID-19 lung infection segmentation methods with limited labeled data has become an urgent need especially in the current situation.

Currently, there are several public non-COVID lung lesion datasets due to other clinical practices, such as MSD Lung for segmentation of lung tumor and NSCLC Pleural Effusion for segmentation of pleural effusion. These non-COVID datasets may serve as potential profit for generalizing useful information to the related COVID-19 infection segmentation task. Wang \textit{et al.} \cite{wang2020does} have proven that pre-training on non-COVID datasets can improve the segmentation performance of COVID-19 infection segmentation. However, the improvement of transfer learning is not stable when encountering large domain difference between datasets, and shared knowledge between COVID-19 and non-COVID lung lesions cannot be fully exploited.

To address these challenges, we propose a novel relation-driven collaborative learning model for annotation-efficient COVID-19 CT lung infection segmentation by exploiting shared knowledge from non-COVID lesions. The network consists of encoders with the same architecture and a shared decoder. The general encoder is adopted to capture general lung lesion features based on multiple non-COVID lesions, while the target encoder is adopted to focus on task-specific features of COVID-19 infections. Features extracted from the two parallel encoders are concatenated for the subsequent decoder part. 
To exploit shared knowledge between COVID and non-COVID lesions, we develop a collaborative learning scheme to regularize the relation consistency between extracted features of given input. 
Our method can enforce the consistency of feature relation among extracted features and encourage the model to explore semantic information from both COVID-19 and non-COVID cases. Besides, the scheme can also be extended to utilize unlabeled COVID-19 data for feature relation regularization and achieve more consistent and robust learning.
The contributions of this work are summarized as follows:

\begin{itemize}
	\item We propose a novel relation-driven collaborative learning model for annotation-efficient segmentation of COVID-19 lung infections from CT images by leveraging shared knowledge from non-COVID lesions to improve the segmentation performance of COVID-19 infections with limited training data.
	\item We present a collaborative learning scheme to explore general semantic information from both COVID-19 and non-COVID cases by regularizing feature-level relation consistency of given input, so as to encourage the model to learn more general and discriminative representation of COVID-19 infections for better segmentation performance. The scheme can also be extended to utilize unlabeled COVID-19 data for the regularization to achieve more consistent and robust learning.
	\item We have conducted extensive experiments on two COVID-19 datasets and two non-COVID lung lesion datasets for 2D and 3D segmentation tasks. The results show that our method achieves superior segmentation performance compared with other methods in the absence of sufficient high-quality COVID-19 data. 
	
\end{itemize}

\begin{figure*}[t]
	\includegraphics[width=18cm]{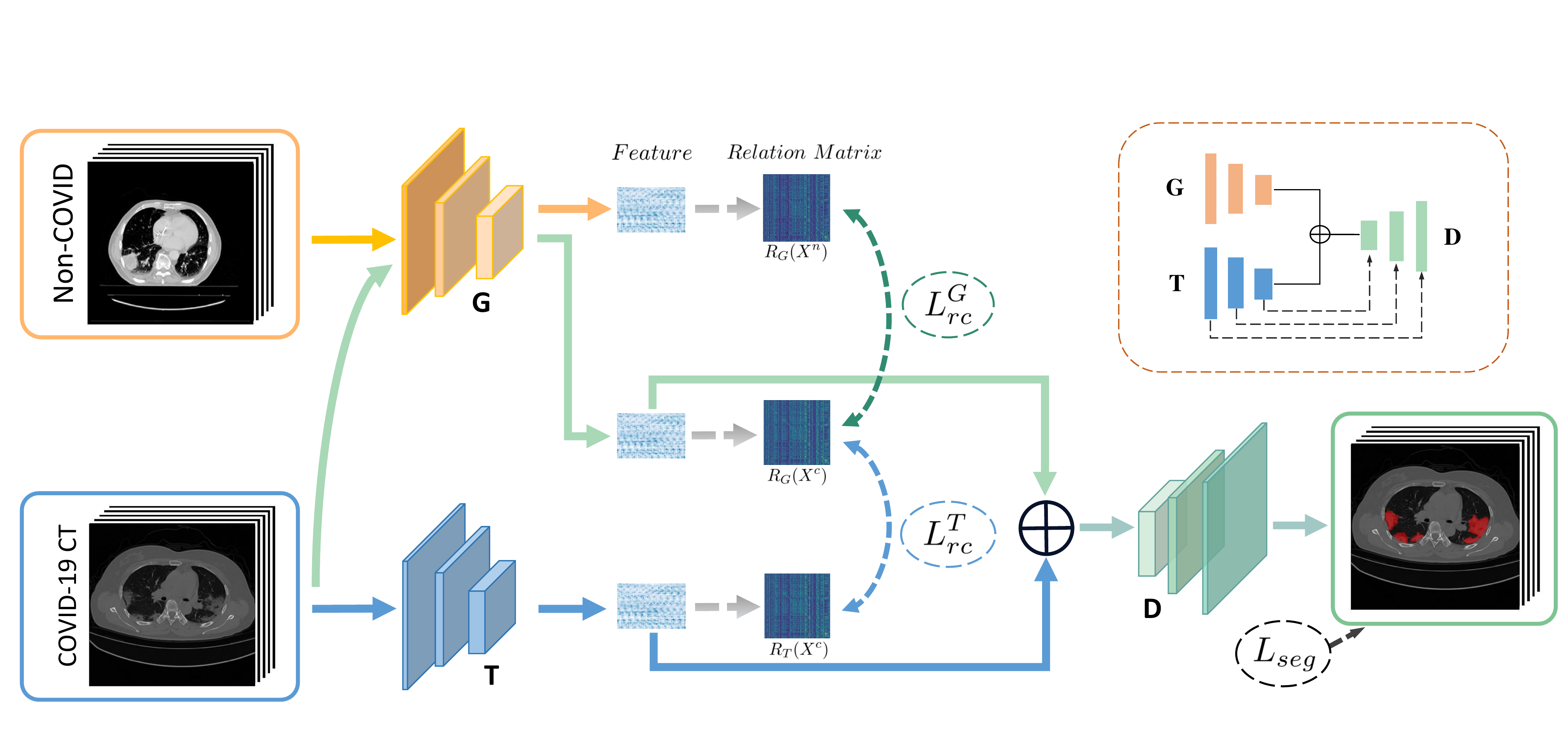}
	\centering
	\caption{The overview of our proposed relation-driven collaborative learning model, where green and blue represent the data flow of general encoder and target encoder for COVID-19 infection segmentation, respectively. Extracted features from these two parallel encoders are concatenated for the input of the shared decoder. To exploit shared knowledge from non-COVID-cases, an additional data flow in orange is adopted. By regularizing feature-level relation consistency of given input, the model is encouraged to explore semantic information from both COVID-19 and non-COVID cases. Since the general encoder is applied to utilize non-COVID data and assist in the learning of COVID-19 segmentation as an auxiliary branch, we only employ the skip connections between the target encoder and shared decoder for the fusion of multi-scale features as shown in the top right corner.}
	\label{architecture}
\end{figure*}

\section{Related Work}

In this section, we briefly review the research related to our work. We first review works on annotation-efficient deep learning for medical image segmentation. Then we review existing works on COVID-19 segmentation and transfer learning approaches for COVID-19.

\subsection{Annotation-efficient Deep Learning}

Compared with natural images, the annotations of medical images are much harder and more expensive to acquire due to following problems: 1) annotating medical images heavily relies on professional diagnosis knowledge of radiologists; 2) most modalities of medical images like CT are 3D volumes, which will take much more time and labor for annotation. To alleviate annotation scarcity, annotation-efficient methods have received great attention in medical image analysis community \cite{cheplygina2019not,tajbakhsh2020embracing}. For example, semi-supervised learning aims at learning from a limited amount of labeled data and a large amount of unlabeled data, which is an effective way to explore knowledge from the unlabeled data \cite{van2020survey}. Weakly supervised learning explores the use of weak annotations like noisy annotations and sparse annotations \cite{ji2019scribble}. Besides, some approaches also aim at integrating multiple related datasets to learn general knowledge \cite{huang2020multi,zhang2020dodnet}. To issue the problem of limited labeled COVID-19 data, in this work, we aim at utilizing existing non-COVID lung lesion datasets for generalizing useful information to related COVID-19 task, so as to achieve better segmentation performance with limited in-domain training data.

\subsection{Research on COVID-19 Segmentation}

Automatic segmentation of COVID-19 infections from CT images is a crucial step to for quantification of the disease progression. Recently, several approaches have been proposed for COVID-19 lung infection segmentation. Shan \textit{et al.} \cite{shan2020lung} propose a deep learning-based system for automatic segmentation and quantification of infection regions. Amyar \textit{et al.} \cite{amyar2020multi} propose to improve the segmentation performance with a multi-task learning approach. Xie \textit{et al.} \cite{xie2020relational} propose a relational approach to leverages structured relationships by introducing a novel non-local neural network module to learn both visual and geometric relationships. Zhou \textit{et al.} \cite{zhou2020automatic} propose a U-Net based segmentation network to incorporate spatial and channel attention for better feature representation.
Other than fully supervised learning, Zheng \textit{et al.} \cite{zheng2020deep} develop a weakly-supervised approach to investigate the potential for automatic detection of COVID-19 based on patient-level label. Fan \textit{et al.} \cite{fan2020inf} present a lung infection segmentation network for 2D CT slices with semi-supervised strategy. Wang \textit{et al.} \cite{wang2020noise} propose a noise-robust framework to learn from noisy labels for the pneumonia lesion segmentation task. Yao \textit{et al.} \cite{yao2020label} use a set of operations to synthesize lesion-like appearances for label-free segmentation. Ma \textit{et al.} \cite{ma2020active} propose an active contour regularized framework using region-scalable fitting to regularize and refine the pseudo labels for semi-supervised infection segmentation.

\subsection{Transfer Learning Approaches for COVID-19}

Transfer learning aims to leverage knowledge and latent features from other datasets by pre-training models on large datasets and fine-tuning trained models on downstream tasks. Due to the problem of limited COVID-19 data, several transfer learning methods have been applied. For example, Chouhan \textit{et al.} \cite{chouhan2020novel} propose an ensemble model to combine outputs from five pre-trained models based on ImageNet. Majeed \textit{et al.} \cite{majeed2020covid} adopt transfer learning procedure and propose a simple CNN architecture with a small number of parameters to distinguish COVID-19 from normal X-rays. Misra \textit{et al.} \cite{misra2020multi} propose a multi-channel pre-trained ResNet architecture to facilitate the diagnosis of COVID-19. For segmentation of COVID-19 infections, Wang \textit{et al.} \cite{wang2020does} evaluate different transfer learning methods and revealed the benefits of transferring knowledge from non-COVID lung lesions. However, transfer learning only takes the advantage of existing models, and non-COVID cases are not utilized in the training procedure of downstream COVID-19 segmentation tasks. Different from these existing methods, our method aims at learning from COVID-19 and non-COVID lung lesions collaboratively to exploit shared semantic information.

\section{Method}

In this section, we first introduce the overview of our proposed method. Then we provide details of our relation-driven collaborative learning scheme and the overall training procedure.

\subsection{Overview}

An overview of our proposed framework is shown in Fig\ref{architecture}. Following the design of standard U-Net \cite{ronneberger2015u,cciccek20163d}, our network consists of two encoders with the same architecture and a shared decoder. Since the encoder serves as a contraction to extract image contextual features, the upper one named general encoder (G) is adopted to capture general lung lesion features based on multiple non-COVID lesions, and the lower one named target encoder (T) is adopted to focus on task-specific features of the target COVID-19 infection segmentation task. 
After that, extracted features from these two parallel encoders are concatenated together for the input of the decoder. The shared decoder (D) serves as a symmetric expanding path to recover the spatial information of the extracted features.
Since our motivation is to use non-COVID lesions to assist in the segmentation of COVID-19 infections, the general encoder can be seemed as an auxiliary branch to extract shared knowledge. Therefore, we only employ the skip connections between the target encoder and shared decoder for the fusion of multi-scale features as shown in the top right corner of Fig\ref{architecture}.

Given a set of samples $\{X_{i}^{c},Y_{i}^{c}\}_{i=1}^{N_{c}}$ from COVID-19 datasets $D_{c}$ and a set of samples $\{X_{i}^{n},Y_{i}^{n}\}_{i=1}^{N_{n}}$ from non-COVID datasets $D_{n}$, where $X$ and $Y$ denote the CT image and corresponding annotation of lung lesions. 
For the segmentation workflow, the general encoder is applied to extract general features, while the target encoder is applied to extract the task-specific features. These extracted features are concatenated and then fed into the decoder part to get the final segmentation results. 
To issue the problem of limited COVID-19 training data, instead of transferring pre-trained models to the downstream learning task of $X^{c}$, we aim to involve $X^{n}$ collaboratively in the training procedure of COVID-19 to exploit shared knowledge from non-COVID cases, which can be used as a guidance for the learning of target COVID-19 infection segmentation. Specifically, a relation-driven collaborative learning scheme is applied to regularize the relation consistency between extracted features of given input and encourage the model to explore semantic information.

\subsection{Relation-driven Collaborative Learning}

Inspired by recent study on data-level regularization with sample relation \cite{liu2020semi}, to exploit shared knowledge between non-COVID and COVID-19 cases for collaborative learning, we propose to regularize feature-level relation consistency of given input, so as to facilitate the learning procedure of COVID-19 lung infections segmentation. 
Based on the assumption that general encoder is adopted to capture general features of lung lesions, and the target encoder is adopted to focus on task-specific features of COVID-19 infection, we aim to enforce the relation of features extracted from these two encoders as guidance for the collaborative learning approach.

To estimate the relation of extracted features, we model the feature relation with channel-wise Gram Matrix \cite{gatys2016neural}. For each input batch with $B$ samples, we average the features within each batch to get the mean representation. We denote the extracted feature maps of encoder as $F \in R ^{C\times H \times W \times D}$ for 3D segmentation networks or $F \in R ^{C\times H \times W}$ for 2D segmentation networks, where $H$, $W$ and $D$ represent the spatial dimension of feature maps, and $C$ represents the channel number. 
To obtain channel-wise feature relation, we reshape the feature maps into $A \in R ^{C\times HWD}$ or $A \in R ^{C\times HW}$. After that, we get the channel-wise Gram Matrix as follows:

\begin{equation}
G = A \cdot A^T
\label{gram_matrix}
\end{equation}
where the value of $m_{th}$ row and $n_{th}$ column $G_{mn}$ is the inner product between the vectorized activation maps $A_{(m)}$ and $A_{(n)}$, representing the similarity between the $m_{th}$ and $n_{th}$ channel of extracted features. Therefore, the final feature relation matrix $R$ is obtained by conducting the $L_{2}$ normalization for each row of $G$ as follows

\begin{equation}
R = [\frac{G_1}{\left\|G_1\right\|_2},...,\frac{G_C}{\left\|G_C\right\|}_2]^T
\label{relation_matrix}
\end{equation}

After the modeling of feature relation, our method encourages the network to learn more general and discriminative representation of COVID-19 infections by regularizing the feature relation consistency among given input, so as to explore semantic information from both COVID-19 and non-COVID cases.

For explicit learning, the network is optimized based on the supervised segmentation loss $\mathcal{L}_{seg}$ between output $\hat{Y}^{c}$ and corresponding ground truth $Y^{c}$. We use the combination of dice loss $\mathcal{L}_{dice}$ and cross entropy loss $\mathcal{L}_{ce}$ as the supervised segmentation loss, and deep supervision \cite{dou20163d} is applied to obtain multi-scale supervision at different scales. The supervised segmentation loss can be summarized as

\begin{equation}
\mathcal{L}_{seg} = \mathcal{L}_{dice}(\hat{Y}^{c} ,Y^{c}) + \mathcal{L}_{ce}(\hat{Y}^{c} ,Y^{c})
\label{loss_seg}
\end{equation}

Besides, to utilize the feature relation for collaborative learning, the non-COVID cases are additionally fed into general encoder to explore the general feature representation and its corresponding feature relation matrix $R_{G}(X^{n})$. To ensure that general encoder can capture general features of lung lesions, our proposed scheme requires the generated feature relation matrices of general encoder to be stable using general relation consistency loss $\mathcal{L}_{rc}^{G}$ defined as

\begin{equation}
\mathcal{L}_{rc}^{G} =  \underset{\{{X^{n},X^{c}}\} \in \{S_{n},S_{c}\} }{\sum}    \lambda_{G} \textbf{ }    \lVert R_{G}(X^{n}) - R_{G}(X^{c}) \rVert ^{2}  
\label{loss_rc_G}
\end{equation}

While for target encoder, 
we enforce the extracted relation matrices of task-specific features to be more discriminative compared with general encoder using target relation consistency loss $\mathcal{L}_{rc}^{T}$ defined as

\begin{equation}
\mathcal{L}_{rc}^{T} =  \underset{\{{X^{n},X^{c}}\} \in \{S_{n},S_{c}\} }{\sum}   - \lambda_{T} \textbf{ }   \lVert R_{G}(X^{c}) - R_{T}(X^{c}) \rVert ^{2}
\label{loss_rc_T}
\end{equation}
where $R_{G}(X^{c})$ and $R_{T}(X^{c})$ denote the feature relation matrices of COVID-19 cases extracted from general encoder and target encoder, respectively. $\lambda_{G}$ and $\lambda_{T}$ are ramp-up weighting coefficients that control the trade-off between the segmentation loss and consistency loss, so as to mitigate the disturbance of consistency loss at early training stage.
Since the network is supervised by limited COVID-19 cases, the training may become unstable and with poor generalization ability. By minimizing feature relation consistency losses $\mathcal{L}_{rc}^{G}$ and $\mathcal{L}_{rc}^{T}$ during the training procedure, the general encoder and target encoder can be enhanced to capture more general and discriminative representation, thereby exploring useful shared knowledge from adequate non-COVID data for better segmentation performance. 

\renewcommand{\algorithmicrequire}{ \textbf{Input:}}     %Use Input in the format of Algorithm
\renewcommand{\algorithmicensure}{ \textbf{Output:}}    %Use Output in the format of Algorithm
\begin{algorithm}[t]
	\caption{Training procedure of our proposed framework.}
	\label{Algorithm1}
	\begin{algorithmic}[1]
		\REQUIRE{A batch of ($X^{c}$,$Y^{c}$) from COVID-19 dataset $D_{c}$ and ($X^{n}$,$Y^{n}$) from non-COVID dataset $D_{n}$.}
		\ENSURE{Trained network $\mathcal{N}$ with parameters $\theta_{G}$,$\theta_{T}$,$\theta_{D}$}
		\WHILE{not converge}
		\STATE ($X^{c}$,$Y^{c}$), ($X^{n}$,$Y^{n}$) $\leftarrow$ sampled from $D_{c}$ and $D_{n}$
		\STATE Generate features of general encoder $F_{G}(X^{c})$ and target encoder $F_{T}(X^{c})$
		\STATE Generate general feature representation $F_{G}(X^{n})$ 
		\STATE Calculate feature relation matrices $R_{G}(X^{c})$, $R_{T}(X^{c})$ and $R_{G}(X^{n})$ as Eq. (1) and (2)
		\STATE Generate segmentation output $\hat{Y}^{c}$
		\STATE Calculate segmentation loss $\mathcal{L}_{seg}$ as Eq. (3)
		\STATE Calculate consistency losses $\mathcal{L}_{rc}^{G}$, $\mathcal{L}_{rc}^{T}$ as Eq. (4) and (5)
		\STATE Update $\theta_{G}$ $\stackrel{+}{\longleftarrow}$ $-\Delta_{\theta_{G}} \mathcal{L}_{rc}^{G}$
		\STATE Update $\theta_{T}$ $\stackrel{+}{\longleftarrow}$ $-\Delta_{\theta_{T}} (\mathcal{L}_{seg} + \mathcal{L}_{rc}^{T})$
		\STATE Update $\theta_{D}$ $\stackrel{+}{\longleftarrow}$ $-\Delta_{\theta_{D}} \mathcal{L}_{seg}$
		\STATE Ramp up the weighting coefficients $\lambda_{G}$ and $\lambda_{T}$
		\ENDWHILE 
		\RETURN {Trained network $\mathcal{N}$}
	\end{algorithmic}
\end{algorithm}

\subsection{Overall Training Procedure}

Algorithm\ref{Algorithm1} presents the detailed training procedure of our framework. 
For the optimation of the network, we update the target encoder and decoder based on the supervised segmentation loss $L_{seg}$. Besides, relation consistency losses $L_{rc}^{G}$ and $L_{rc}^{T}$ are used to update the general encoder and target encoder, respectively. The collaborative learning scheme allow the two parallel encoders to benefit from each other’s guidance, encouraging the model to explore semantic information from both COVID-19 and non-COVID cases.

\subsection{Feature Relation Regularization with Unlabeled Data}

Since our proposed general relation consistency loss $\mathcal{L}_{rc}^{G}$ and target relation consistency loss $\mathcal{L}_{rc}^{T}$ do not require segmentation label, our proposed method can be straightforwardly extended to utilize unlabeled COVID-19 data for feature relation regularization.
Specifically, we only activate the supervised segmentation loss $\mathcal{L}_{seg}$ for labeled data, while computing the relation consistency losses $\mathcal{L}_{rc}^{G}$ and $\mathcal{L}_{rc}^{T}$ for all the training data.
In this way, unlabeled data can be leveraged for the regularization to achieve more consistent and robust learning.

\begin{table*}[t]
	\centering
	\caption{Details of 3D U-Net architecture used in our experiments. Note that the general encoder and target encoder are with the same architecture as shown in the left column.}
	\label{Table1}
	\renewcommand\arraystretch{1.5}
	\begin{tabular}{p{2cm}|p{1.5cm}p{4cm}|p{1.5cm}p{4cm}}
		\hline
		feature size  & \multicolumn{2}{c}{Encoder (G / T)}  & \multicolumn{2}{c}{Decoder (D)}           \\ \hline
		1x56x160x192  & input &                             & output & conv(1x1x1)-sigmoid             \\
		32x56x160x192 & conv1 & conv(1x3x3)-IN-LReLU        & conv10 & conv(1x3x3)-IN-LReLU            \\
		64x56x80x96   & down1 & strided conv(1,2,2)         & up10   & transposed conv(1,2,2) - conv1  \\
		64x56x80x96   & conv2 & conv(3x3x3)-IN-LReLU        & conv9  & conv(3x3x3)-IN-LReLU            \\
		128x28x40x48  & down2 & strided conv(2,2,2)         & up9    & transposed conv(2,2,2) - conv2  \\
		128x28x40x48  & conv3 & conv(3x3x3)-IN-LReLU        & conv8  & conv(3x3x3)-IN-LReLU            \\
		256x14x20x24  & down3 & strided conv(2,2,2)         & up8    & transposed conv(2,2,2) - conv3  \\
		256x14x20x24  & conv4 & conv(3x3x3)-IN-LReLU        & conv7  & conv(3x3x3)-IN-LReLU            \\
		320x7x10x12   & down4 & strided conv(2,2,2)         & up7    & transposed conv(2,2,2) - conv4  \\
		320x7x10x12   & conv5 & conv(3x3x3)-IN-LReLU        & conv6  & conv(3x3x3)-IN-LReLU            \\
		320x7x5x6     & down5 & strided conv(1,2,2)         & up6    & transposed conv(1,2,2) - conv5  \\ \hline
	\end{tabular}
\end{table*}

\begin{table*}[t]
	\centering
	\caption{Details of 2D U-Net architecture used in our experiments. Note that the general encoder and target encoder are with the same architecture as shown in the left column.}
	\label{Table11}
	\renewcommand\arraystretch{1.5}
	\begin{tabular}{p{2cm}|p{1.5cm}p{3cm}|p{1.5cm}p{4cm}}
		\hline
		feature size  & \multicolumn{2}{c}{Encoder (G / T)}  & \multicolumn{2}{c}{Decoder (D)}           \\ \hline
		1x448x384     & input &                             & output & conv(1x1)-sigmoid             \\
		32x448x384    & conv1 & conv(3x3)-IN-LReLU        & conv12 & conv(3x3)-IN-LReLU            \\
		64x224x192    & down1 & strided conv(2,2)         & up12   & transposed conv(2,2) - conv1  \\
		64x224x192    & conv2 & conv(3x3)-IN-LReLU        & conv11 & conv(3x3)-IN-LReLU            \\
		128x112x96    & down2 & strided conv(2,2)         & up11   & transposed conv(2,2) - conv2  \\
		128x112x96    & conv3 & conv(3x3)-IN-LReLU        & conv10 & conv(3x3)-IN-LReLU            \\
		256x56x48     & down3 & strided conv(2,2)         & up10   & transposed conv(2,2) - conv3  \\
		256x56x48     & conv4 & conv(3x3)-IN-LReLU        & conv9  & conv(3x3)-IN-LReLU            \\
		480x28x24     & down4 & strided conv(2,2)         & up9    & transposed conv(2,2) - conv4  \\
		480x28x24     & conv5 & conv(3x3)-IN-LReLU        & conv8  & conv(3x3)-IN-LReLU            \\
		480x14x12     & down5 & strided conv(2,2)         & up8    & transposed conv(2,2) - conv5  \\
		480x14x12     & conv6 & conv(3x3)-IN-LReLU        & conv7  & conv(3x3)-IN-LReLU            \\
		480x7x6       & down6 & strided conv(2,2)         & up7    & transposed conv(2,2) - conv6  \\ \hline
	\end{tabular}
\end{table*}

\section{Experiments}

\subsection{Dataset Introduction}

\subsubsection{COVID-19 Dataset}
We select out two public COVID-19 lung infection segmentation datasets for our experiments.
The first dataset contains 20 CT volumes with over 1800 annotated slices released by Coronacases Initiative and Radiopaedial, which is publicly available at \footnote{https://zenodo.org/record/3757476\#.X4ABeYvivid}. The annotation of infections is labeled by two radiologists and verified by an experienced radiologist by Ma \textit{et al.} \cite{ma2020towards}.
The second dataset is COVID-19 CT Segmentation dataset \footnote{http://medicalsegmentation.com/covid19/} collected by the Italian Society of Medical and Interventional Radiology, which contains 100 2D axial CT slices from different COVID-19 patients. A radiologist segmented the CT images using different labels for identifying lung infections.

\subsubsection{Non-COVID Lung Lesion Datasets}
In order to explore relevant information from non-COVID lung lesions to promote the annotation-efficient training of COVID-19 cases, we select out two public non-COVID lung lesion segmentation datasets for our following experiments. The first dataset is MSD Lung Tumor Dataset of Medical Segmentation Decathlon (MSD) Challenge \cite{simpson2019large} in MICCAI 2018 \footnote{http://medicaldecathlon.com/}. This dataset is comprised of patients with non-small cell lung cancer from Stanford University (Palo Alto, CA, USA) publicly available through TCIA. The tumor is annotated by an expert thoracic radiologist and 63 labeled CT volumes are used. 
The second dataset is NSCLC Pleural Effusion Dataset \footnote{https://wiki.cancerimagingarchive.net/display/Public/NSCLC-Radiomics}. This dataset contains 78 CT volumes with annotation of pleural effusion. To exploit general features of lung lesions, we combine MSD and NSCLC datasets to form a non-COVID multi-lesion dataset in the following experiments.

\begin{table*}[t]
	\centering
	\caption{Quantitative results of 5-fold cross validation of ablation analysis in our experiments.}
	\label{Table2}
	\renewcommand\arraystretch{1.7}
	\begin{tabular}{c|cccccc|cccccc}
		\hline \hline
		{\multirow{2}{*}{Method}} & \multicolumn{6}{c}{DSC}     & \multicolumn{6}{c}{NSD}      \\ \cline{2-13}
		{}                        & Fold 0 & Fold 1 & Fold 2 & Fold 3 & Fold 4 & Avg   & Fold 0 & Fold 1 & Fold 2 & Fold 3 & Fold 4 & Avg \\ \hline
		3D nnUNet                 & 0.681  & 0.713  & 0.662  & 0.681  & 0.627  & 0.673$\pm$0.223 & 0.709  & 0.718  & 0.717  & 0.708  & 0.649 & 0.700$\pm$0.224 \\ \hline
		Ours (baseline)           & 0.689  & 0.721  & 0.712  & 0.720  & 0.632  & 0.695$\pm$0.205 & 0.709  & 0.747  & 0.770  & 0.764  & 0.649 & 0.728$\pm$0.216 \\ 
		Ours ($L_{rc}^{G}$ )      & 0.701  & 0.727  & 0.727  & 0.710  & 0.625  & 0.699$\pm$0.210 & 0.747  & 0.758  & 0.790  & 0.763  & 0.648 & 0.740$\pm$0.221 \\ 
		Ours ($L_{rc}^{T}$ )      & 0.696  & 0.725  & 0.729  & 0.699  & 0.638  & 0.697$\pm$0.204 &  0.744  & 0.756  & 0.796  & 0.752 & 0.654 & 0.739$\pm$0.213 \\ 
		Ours ($L_{rc}^{G}$+$L_{rc}^{T}$)  & 0.723  & 0.728  & 0.718  & 0.720  & 0.625  & \textbf{0.703$\pm$0.193} & 0.756  & 0.760  & 0.790  & 0.771  & 0.631 & \textbf{0.742$\pm$0.203} \\	  \hline \hline
	\end{tabular}
\end{table*}

\begin{figure*}[t]
	\includegraphics[width=18cm]{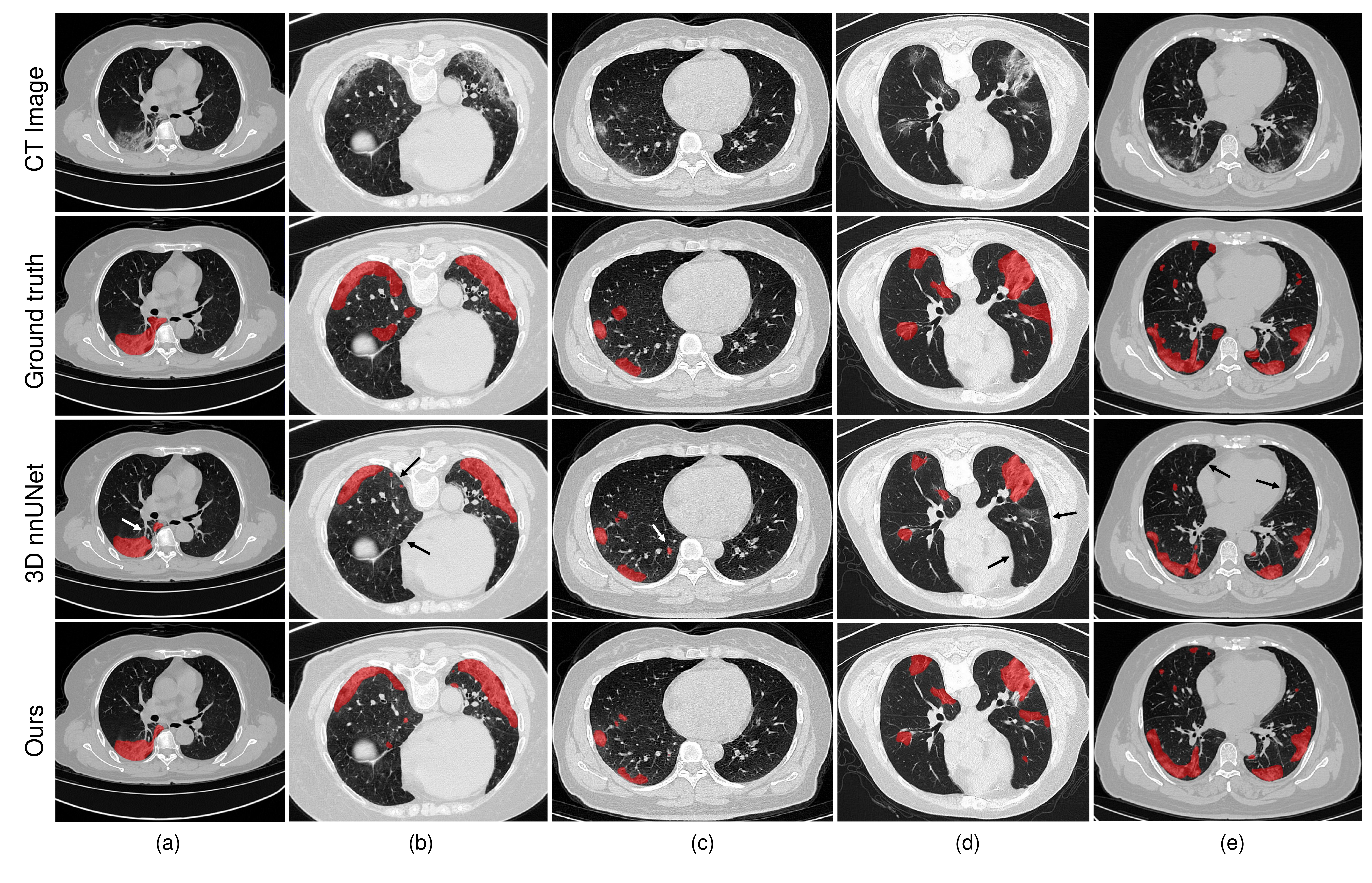}
	\centering
	\caption{Visual comparison of COVID-19 infection segmentation by different methods on 3D COVID-19 segmentation benchmark dataset from axial view. }
	\label{fig5}
\end{figure*}

\subsection{Experimental Settings}

\subsubsection{3D Experiments on CT Volumes}

For 3D experiments of CT Volumes, to make a fair comparison, we follow the task settings of COVID-19 benchmarks in \cite{ma2020towards}. For the COVID-19 dataset, we make the same 5-fold cross validation based on pre-defined dataset split. Each fold contains 4 scans (20\%) for training and 16 scans (80\%) for testing. For non-COVID lung lesion datasets, we randomly select 80\% of the data for training and the rest of 20\% for validation. 

We use 3D U-Net \cite{cciccek20163d} as the backbone network. Details of network architecture is shown in Table \ref{Table1}.
The input patch size is set as 56$\times$160$\times$192 with batch size of 2. Stochastic gradient descent (SGD) optimizer is used for training with initial learning rate of 0.01 and momentum of 0.99.

\subsubsection{2D Experiments on Axial CT Slices}

To compare our method with state-of-the-art methods for 2D medical image segmentation, we make comparison experiments based on 2D COVID-19 CT slices. Following the same task settings of \cite{fan2020inf}, we use the same 50 images for training and validation, and the remaining 50 slices for testing.
For non-COVID lung lesion datasets, we randomly select out 100 2D slices with lung lesions from different CT scans.
Besides, unlabeled training set of COVID-SemiSeg Dataset \cite{fan2020inf} is used to evaluate the effectiveness of our proposed method to utilize unlabeled COVID-19 data.

We use 2D U-Net \cite{ronneberger2015u} as the backbone network. Details of network architecture is shown in Table \ref{Table11}. The input patch size is set as 448$\times$384 with batch size of 2. Stochastic gradient descent (SGD) optimizer is used for training with initial learning rate of 0.01 and momentum of 0.99.

\subsection{Implementation Details and Evaluation Metrics}

All the experiments in our work are implemented in Pytorch \cite{paszke2019pytorch} and trained on NVIDIA Tesla V100 GPUs. 
Our backbone network is based on nnUNet \cite{isensee2020nnu} that achieved state-of-the-art results in 23 segmentation challenges with automatically designed U-Net according to the dataset properties.
To unify the setting for our collaborative learning approach, we use planned network architectures of COVID-19 infection segmentation task for our framework.
Following \cite{laine2016temporal}, we use a Gaussian ramp-up function $ \lambda(t)=0.1*e^{-5(1-T/T_{max})} $ to control the balance between supervised loss and consistency loss, where $T$ represents the current training step and $T_{max}$ represents the maximum training step.

Motivated by the evaluation methods of the medical image segmentation decathlon \cite{simpson2019large}, we employ two complementary metrics to evaluate the segmentation performance. Dice Similarity Coefficient (DSC), a region-based measure is used to measure the region mismatch, and Normalized surface Dice (NSD), a boundary-based measure is used to evaluate how close the segmentation and ground truth surfaces are to each other. 
Both metrics take the values in [0,1] and higher scores represent better segmentation performance. Let $G$ and $S$ denote the ground truth and the segmentation result, respectively. The two metrics are defined as follows:

\begin{equation}
DSC(G, S) = \frac{2|G\cap S|}{|G| + |S|};
\label{DSC}
\end{equation}

\begin{equation}
NSD(G, S) = \frac{|\partial G\cap B_{\partial S}^{(\tau)}| + |\partial S\cap B_{\partial G}^{(\tau)}|}{|\partial G| + |\partial S|}.
\label{NSD}
\end{equation}
where $B_{\partial G}^{(\tau)}, B_{\partial S}^{(\tau)}$ denote the border regions of ground truth and segmentation surface at a threshold $\tau$ to tolerate the inter-rater variability of the annotators. We set $\tau$ = 3mm for the evaluation of segmentation results in the following experiments.
% Other three metrics
Besides, we also consider three other evaluation metrics in 2D experiments. Sensitivity (Sen) denotes the percentage of positive instances correctly identified. Specificity (Spec) denotes the percentage of predicted positive instances that are correctly identified. Mean Absolute Error (MAE) measures the pixel-wise error between segmentation output and corresponding groundtruth.

\begin{table*}[t]
	\caption{Quantitative results of 5-fold cross validation of 3D comparison experiments with state-of-the-art methods.
		The best results are shown in \textcolor{red}{red} font and the second-best results in \textcolor{blue}{blue} font.}
	\label{Table3}
	\renewcommand\arraystretch{1.7}
	\scriptsize
	\begin{tabular}{p{3.2cm}<{\centering}|cccccc|cccccc}
		\hline \hline
		{\multirow{2}{*}{Method}} & \multicolumn{6}{c}{DSC}     & \multicolumn{6}{c}{NSD}    \\ \cline{2-13}
		{}                  & Fold 0 & Fold 1 & Fold 2 & Fold 3 & Fold 4 & Avg   & Fold 0 & Fold 1 & Fold 2 & Fold 3 & Fold 4 & Avg \\ \hline
		3D nnUNet    &0.681  & 0.713  & 0.662  & 0.681  & \textcolor{red}{0.627}  & 0.673$\pm$0.223 & 0.709  & 0.718  & 0.717  & 0.708  & \textcolor{red}{0.649} & 0.700$\pm$0.224 \\

		Pre-train on MSD    &0.679  & 0.706  & \textcolor{red}{0.724} & 0.708  & 0.623  & 0.688$\pm$0.201 & 0.706  & 0.708  & \textcolor{blue}{0.785} & 0.724  & \textcolor{blue}{0.642} & 0.713$\pm$0.225 \\
		
		Pre-train on NSCLC  &0.696  & 0.716  & 0.673  & 0.690  & 0.579  & 0.671$\pm$0.228 & 0.714  & 0.720  & 0.734  & 0.707  & 0.590 & 0.693$\pm$0.248 \\

		Pre-train on Multi-lesion$^{*}$  &0.702  & \textcolor{red}{0.736}  & 0.703  & \textcolor{blue}{0.725}  & 0.612  & 0.696$\pm$0.213 & 0.715  & 0.730  & 0.755  & 0.754  & 0.628 & 0.717$\pm$0.227 \\
		
		Multi-encoder$^{*}$   &\textcolor{blue}{0.712}  & \textcolor{blue}{0.732}  & \textcolor{blue}{0.721} & \textcolor{red}{0.742}  & 0.608  & \textcolor{blue}{0.703$\pm$0.201} & \textcolor{blue}{0.735}  & \textcolor{blue}{0.740}  & \textcolor{blue}{0.785}  & \textcolor{red}{0.772}  & 0.629 & \textcolor{blue}{0.732$\pm$0.218} \\  \hline
		\textbf{Ours-3D}         &\textcolor{red}{0.723}  & 0.728 & 0.718 & 0.720  & \textcolor{blue}{0.625} & \textcolor{red}{0.703$\pm$0.193} & \textcolor{red}{0.756}  &  \textcolor{red}{0.760} & \textcolor{red}{0.790} & \textcolor{blue}{0.771} & 0.631 & \textcolor{red}{0.742$\pm$0.203}  \\ 	\hline \hline
	\end{tabular}
	~\\ ~\\ Note:   $\quad$ $*$ denotes the results from \cite{wang2020does} where additional non-COVID datasets are used.
\end{table*}

\subsection{Ablation Analysis}

To evaluate the effectiveness of the key components in our framework, we conduct ablation studies by removing the feature relation consistency loss. As shown in Table \ref{Table2}, it is observed that all our methods can achieve better performance on all metrics compared with fully supervised methods, showing the effectiveness of our method. Besides, the usage of $L_{rc}^{G}$ and $L_{rc}^{T}$ can both further improve the segmentation performance compared with baseline.
When removing the target relation consistency, the average segmentation performance of five folds is degraded by 0.4\% and 0.2\% on DSC and NSD, respectively. The result proves that the usage of target relation consistency loss $L_{rc}^{T}$ can enforce the target encoder to be more discriminative, so as to improve the segmentation performance. However, the improvement is susceptible to the domain difference. 
Besides, we also conduct experiments of our backbone by removing the general relation consistency loss $L_{rc}^{G}$. In this way, the general encoder is frozen and are not updated during the training procedure, which means that the knowledge transfer is not available. 
The experimental results demonstrate that the average segmentation performance is degraded by 0.6\% and 0.3\% on DSC and NSD, showing the importance of knowledge transfer in our collaborative learning scheme.
Some segmentation results of our method and 3D nnUNet are illustrated in Fig.\ref{fig5} for visual comparison. 
As shown in the figure, our method can generate segmentation results with more accurate boundaries in Fig.\ref{fig5} (a)(b), and less segmentation mistakes in small infection areas in Fig.\ref{fig5} (c)(d)(e). These results demonstrate that the collaborative learning approach can better exploit shared knowledge from non-COVID cases, leading to better performance when generalizing on test data.

\begin{figure*}[t]
	\includegraphics[width=18cm]{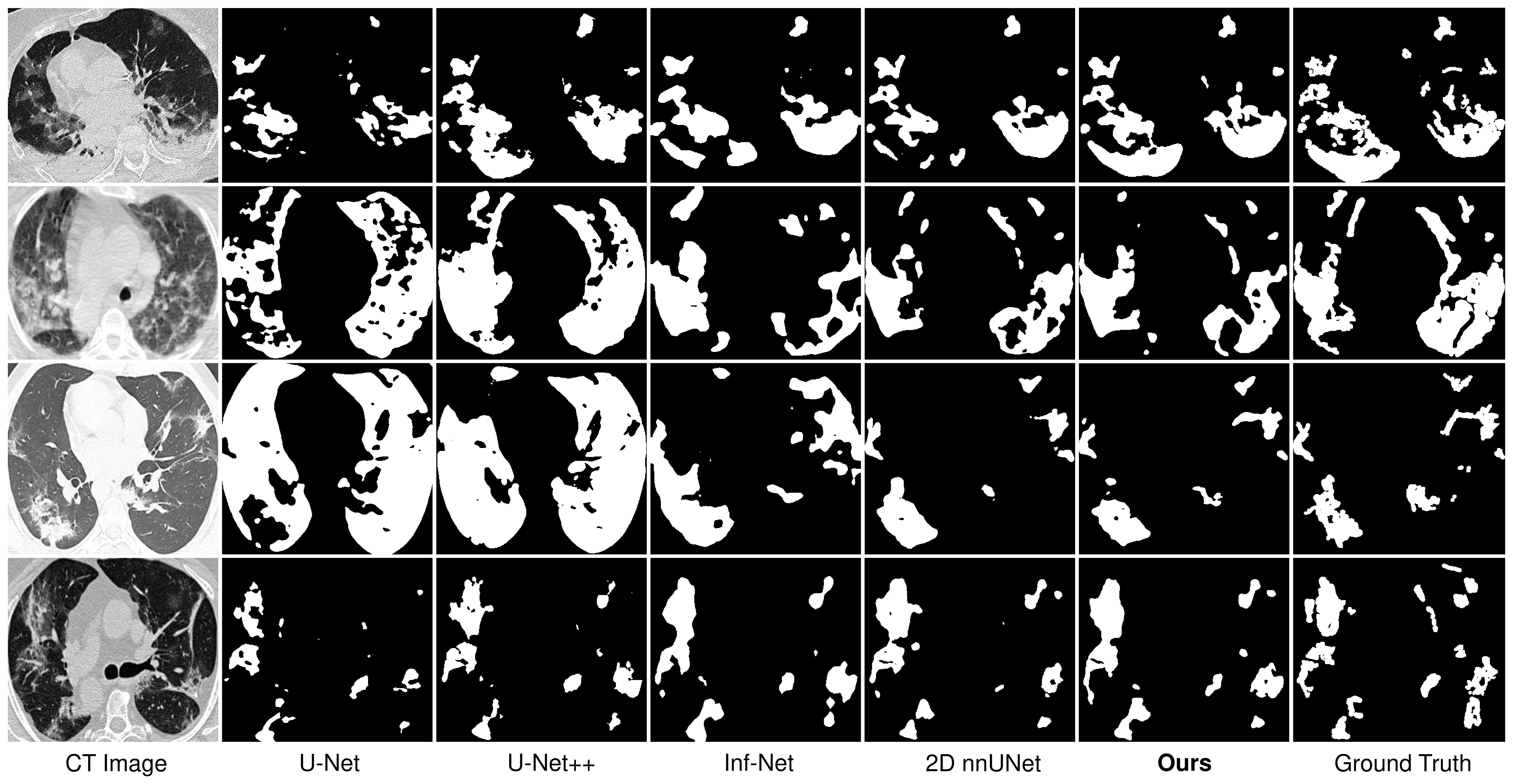}
	\centering
	\caption{Visual comparison of COVID-19 infection segmentation by different methods on 2D COVID-SemiSeg dataset. As can be observed, our method can generate segmentation results with more accurate boundaries and less segmentation mistakes in small infection areas, which is closer to the ground truth.}
	\label{2dresults}
\end{figure*}

\subsection{3D Comparison Experiments on COVID-19 Segmentation Benchmark Dataset}

To demonstrate the effectiveness of our method, we conduct extensive comparison experiments with other state-of-the-art methods. To ensure a fair comparison, all methods are experimented with the same network backbone and experimental settings. Segmentation models trained from scratch with only COVID-19 cases serve as our baseline results. 
Besides, as a simple and intuitive approach, pre-training segmentation models on non-COVID cases and fine-tuning on COVID-19 cases are utilized as comparison
results for learning from both COVID-19 and non-COVID cases. The quantitative experimental results are shown in Table \ref{Table3}. From the results, we can observe that transferring pre-trained models to COVID-19 infection segmentation tasks can generally improve the performance of training from scratch with only COVID-19 cases on most experiments, and using multi-lesion is superior to single-lesion when more general representations can be utilized to help COVID-19 infection tasks, with 2.3\% and 1.7\% improvements in DSC and NSD, respectively. 
However, these transfer learning methods show instability under different data distribution in five-fold cross validation experiments. The rationale is that the transfer ability largely depends on the domain difference between datasets. When there exists a large domain distance between non-COVID and limited COVID-19 training cases, transfer learning may somehow mislead the learning procedure. 

In \cite{wang2020does}, the authors propose a multi-encoder architecture to freeze the non-COVID pre-trained encoder as an additional feature extractor for the training of COVID-19 cases. Features from the frozen adapted-encoder and reinitialized self-encoder are concatenated for the subsequent decoder. However, their workflow is still based on transfer learning, that training a network first on non-COVID cases and then on COVID-19 cases with foregoing pre-trained parameters. The main limitation is that the learning procedures of two tasks are separate. Therefore, the shared knowledge of non-COVID and COVID-19 cases cannot be fully exploited.
It is observed that our method takes advantage of collaborative learning between two encoders and interactively improves the overall learning procedure.
As a consequence, our method achieves higher segmentation performance with an averaged DSC of 70.3\% and averaged NSD of 74.2\%. 
Comparing with training from scratch, exploiting shared knowledge from non-COVID lesions can achieve further improvements with up to 3.0\% in DSC and 4.2\% in NSD.
Paired T-test shows that the improvements are statistically significant at $p < 0.05$, validating the effectiveness of our proposed method.

\begin{table}[!t]
	\caption{Quantitative results of 2D comparison experiments with state-of-the-art methods for fully supervised segmentation and semi-supervised segmentation.}
	\label{Table4}
	\centering
	\scriptsize
	\renewcommand\arraystretch{1.6}
	\begin{tabular}{c|p{1cm}<{\centering}p{1cm}<{\centering}p{1cm}<{\centering}p{1cm}<{\centering}}
		\hline \hline
		Methods & Dice & Sensitivity & Specificity & MAE \\ 	\hline
		U-Net \cite{ronneberger2015u} 	& 0.439 & 0.534 & 0.858 & 0.186 \\
		Attention-UNet \cite{oktay2018attention}  & 0.583 & 0.637 & 0.921 & 0.112\\
		Gated-UNet \cite{schlemper2019attention} 	&  0.623& 0.658 & 0.926 & 0.102 \\
		Dense-UNet \cite{li2018h} 		& 0.515 & 0.594 & 0.840 & 0.184 \\
		U-Net++ \cite{zhou2019unet++} & 0.581  & 0.672 & 0.902 & 0.120 \\
		Inf-Net \cite{fan2020inf} 		& 0.682 & 0.692 & 0.943 & 0.082 \\ 
		Semi-Inf-Net \cite{fan2020inf}	& 0.739 & 0.725 & 0.960 & 0.064 \\
		2D nnUNet \cite{isensee2020nnu} & 0.747 & 0.829 & 0.953 & 0.063 \\ \hline
		\textbf{Ours-2D}          & \textbf{0.765} & \textbf{0.839} & \textbf{0.955} & \textbf{0.056} \\ 
		\textbf{Ours-2D-UN}          & \textbf{0.767} & \textbf{0.839} & \textbf{0.957} & \textbf{0.055} \\  \hline \hline
	\end{tabular}
\end{table}

\subsection{2D Comparison Experiments on COVID-SemiSeg Dataset}

In this subsection, we compare our method with state-of-the-art methods for 2D medical image segmentation, including U-Net \cite{ronneberger2015u}, U-Net++ \cite{zhou2019unet++}, Dense-UNet \cite{li2018h}, Attention-UNet \cite{oktay2018attention}, Gated-UNet \cite{schlemper2019attention}, Inf-Net and Semi-Inf-Net \cite{fan2020inf}.
Quantitative results are shown in Table \ref{Table4}. As can be observed, our proposed method outperforms all comparing methods on all evaluation metrics by a large margin, validating the effectiveness of our framework. Paired T-test shows that the improvements are statistically significant at $p < 0.05$.
Besides, we visualize some segmentation results of our method in Fig.\ref{2dresults}. These results indicate that our segmentation results are closer to the ground truth with less mis-segmented areas and outperform other methods significantly.
For semi-supervised setting, we additionally integrate unlabeled COVID-19 cases of COVID-SemiSeg Dataset into the relation-driven training in our framework. These unlabeled cases can be utilized for the regularization of feature relation to achieve more consistent and robust learning and further improve the segmentation performance slightly.

\subsection{Comparison on Different Datasets for Shared Knowledge Learning}

To better demonstrate the effectiveness of our proposed feature relation-driven learning, we make extensive experiments on several lesion segmentation datasets of medical images with different relation to COVID-19 infections.
In addition to the lung tumor and pleural effusion datasets introduced before, we use LiTS dataset \cite{bilic2019liver} with liver tumor annotations in abdominal CT volumes as non-COVID lesions in our method for comparison.
Besides, to make comparison between intra-disease and inter-disease relations, we use another multi-national CT dataset with labeled ground glass opacities \cite{roth2021rapid} as an out-of-domain dataset for the learning of general branch, which is more relevant with similar appearance to the target dataset in our framework. 

In our experiments, we follow the settings of our 2D experiments with the same network backbone and implementation details.
To make quantitative comparisons, we select out the same amount of 60 cases from these different datasets for the general branch, and 10 cases from the COVID-19 segmentation dataset for target branch.
The experimental results are shown in Table \ref{Table5}.
It can be observed from the table that using intra-disease dataset that are more related to the target dataset can achieve better performance compared with other datasets under the same condition, which proves that more similar appearance can lead to more significant improvement by exploiting shared knowledge. 
Specifically, we observe that using non-lung lesions can also obtain comparable results compared with experiments using non-COVID lung lesions like lung tumor and pleural effusion.

\begin{table}[!t]
	\caption{Quantitative comparison using different datasets as non-COVID cases for shared knowledge learning. All the experiments are performed with the same amount of 60 non-COVID cases and 10 COVID-19 cases under the same settings.}
	\label{Table5}
	\centering
	\scriptsize
	\renewcommand\arraystretch{1.6}
	\begin{tabular}{p{3.8cm}<{\centering}|p{0.6cm}<{\centering}p{0.8cm}<{\centering}p{0.8cm}<{\centering}p{0.6cm}<{\centering}}
		\hline \hline
		Dataset / description      & Dice & Sensitivity & Specificity & MAE \\ 	\hline
		None / COVID only      &  0.693  & 0.774  & 0.941 & 0.080 \\ \hline
		LiTS / liver tumor        & 0.711 & 0.782 & 0.942 & 0.074 \\
		MSD Lung / lung nodules       & 0.712 & 0.764 & \textbf{0.945} & 0.073 \\
		NSCLC / pleural effusion   & 0.702  & 0.781 & 0.941 & 0.072\\ 
		COVID-19-20 / groundglass opacities   & \textbf{0.714} & \textbf{0.792} & 0.942 & \textbf{0.070} \\ \hline \hline
	\end{tabular} 
\end{table}

\subsection{Visual Analysis of Our Method}

\begin{figure}[]
	\includegraphics[width=9cm]{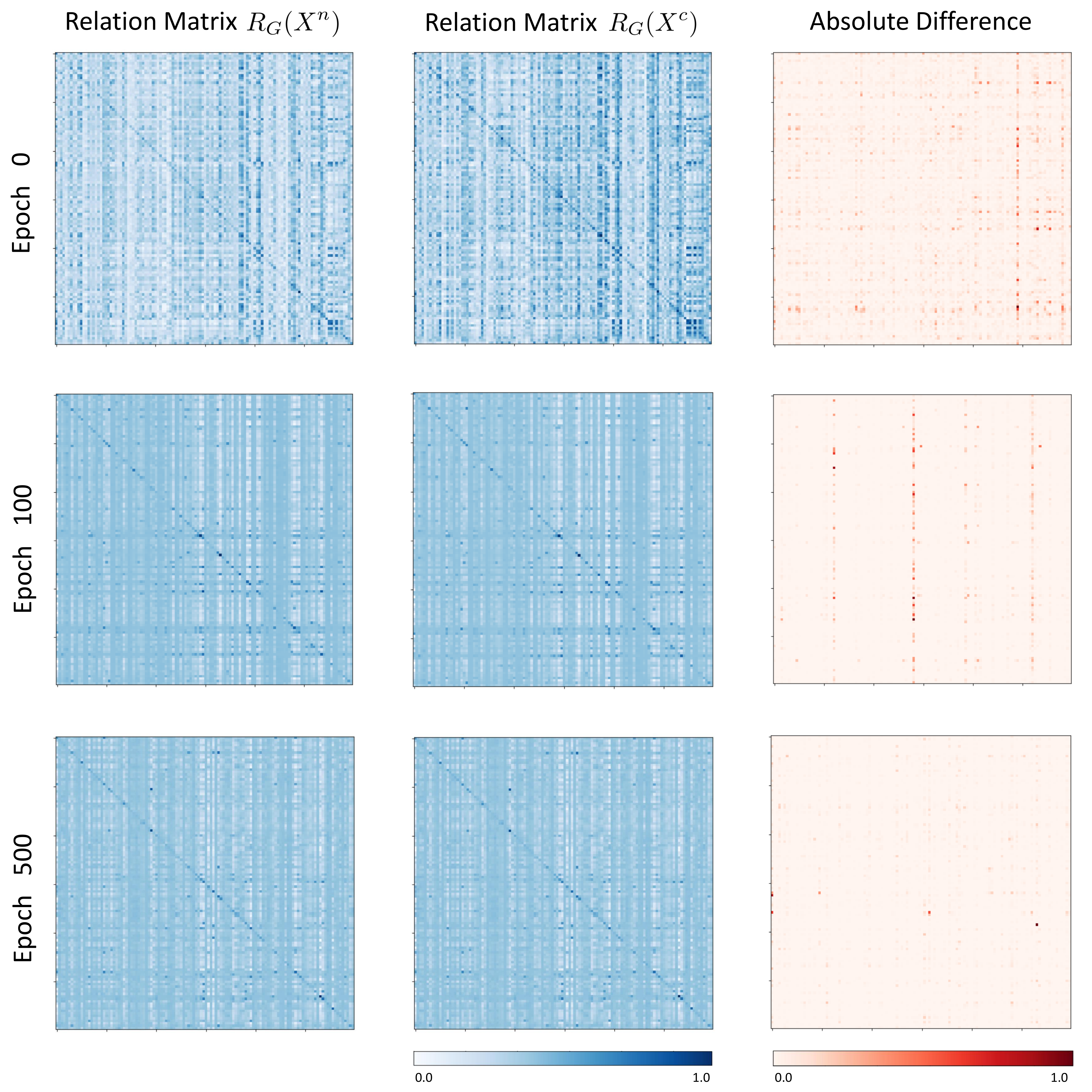}
	\centering
	\caption{Visualization of the general feature relation matrices of non-COVID cases (left column) and COVID-19 cases (middle column) and their absolute difference (right column) during the training procedure. }
	\label{visual1}
\end{figure}

\begin{figure}[]
	\includegraphics[width=9cm]{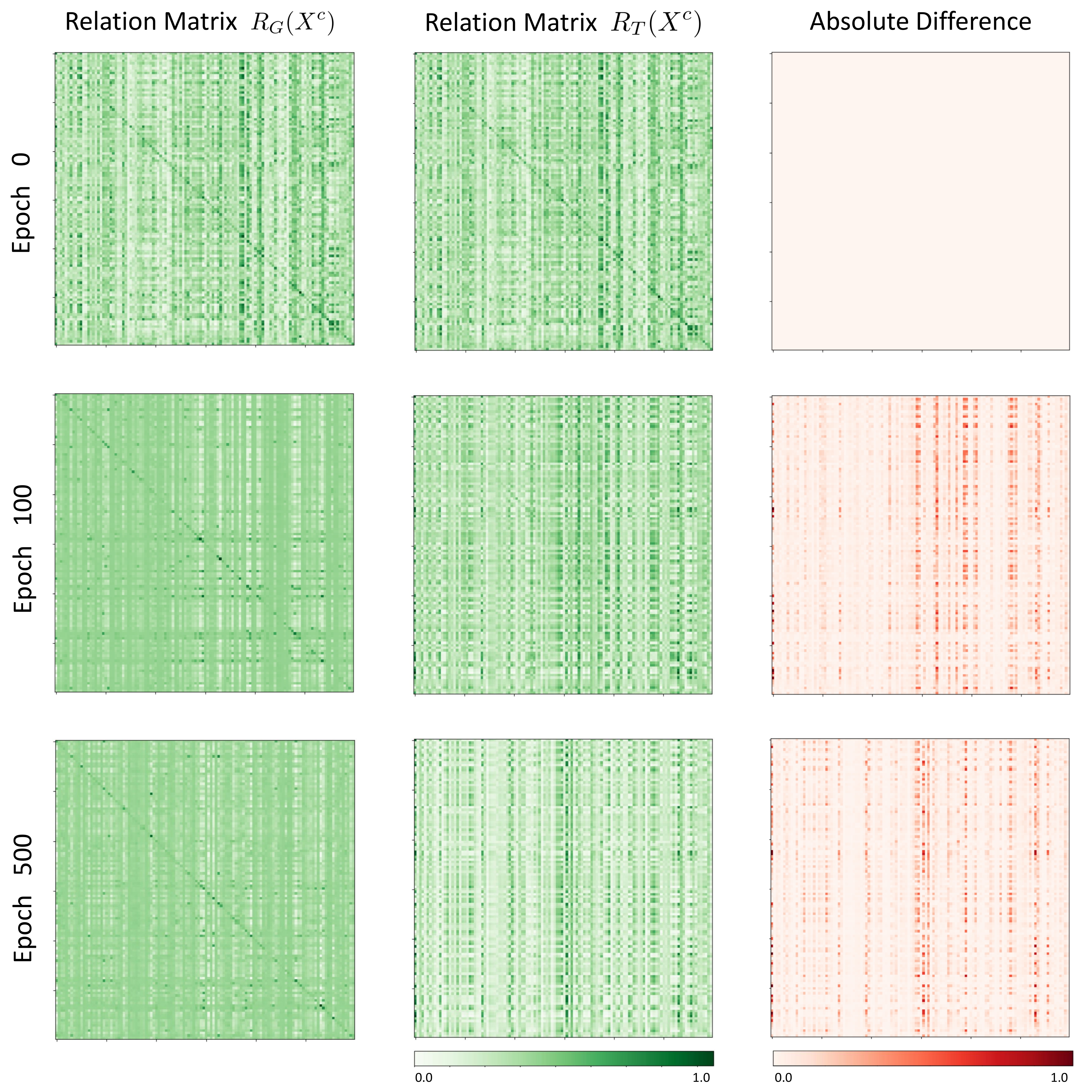}
	\centering
	\caption{Visualization of the general feature relation matrix (left column) and target feature relation matrix (middle column) of COVID-19 cases and their absolute difference (right column) during the training procedure. }
	\label{visual2}
\end{figure}

To visualize learning procedure of our method, we show some examples of general and target feature relation matrices at different epochs during the network training procedure in Fig.\ref{visual1} and Fig.\ref{visual2}. The absolute differences of these two matrices are shown in the right column in red to clearly visualize the alignment of matrices.It can be observed in Fig.\ref{visual1} that as the training goes on, the general encoder gradually produces relation matrices with higher response at the same channel. Meanwhile, the absolute differences of feature relation matrices of non-COVID and COVID-19 cases extracted from general encoder are gradually decreased, indicating that the general encoder learns more general and robust representations of lung lesions.Besides, as observed in Fig.\ref{visual2}, the absolute differences of general and target feature relation matrices of COVID-19 cases are gradually increased and tend to be stable as the training goes on, indicating that the target encoder is gradually enforced to focus on task-specific features and learn more discriminative representations compared with general encoder.

\section{Discussion}

With the outbreak of COVID-19 all over the world, designing effective automated tools for fighting against COVID-19 is highly demanded to improve the efficiency of clinical approaches and reduce the tedious workload of clinicians and radiologists. However, accurate segmentation of COVID-19 lung infections is a challenging task due to the large appearance variance of COVID-19 lesions of patients in different severity level, and existing data-driven segmentation methods mainly rely on large amount of well annotated data. In order to mitigate the insufficiency of labeled COVID-19 CT scans, it is essential and meaningful to develop annotation-efficient segmentation methods for the COVID-19 lung infection segmentation task. 

\begin{figure}[]
	\includegraphics[width=9cm]{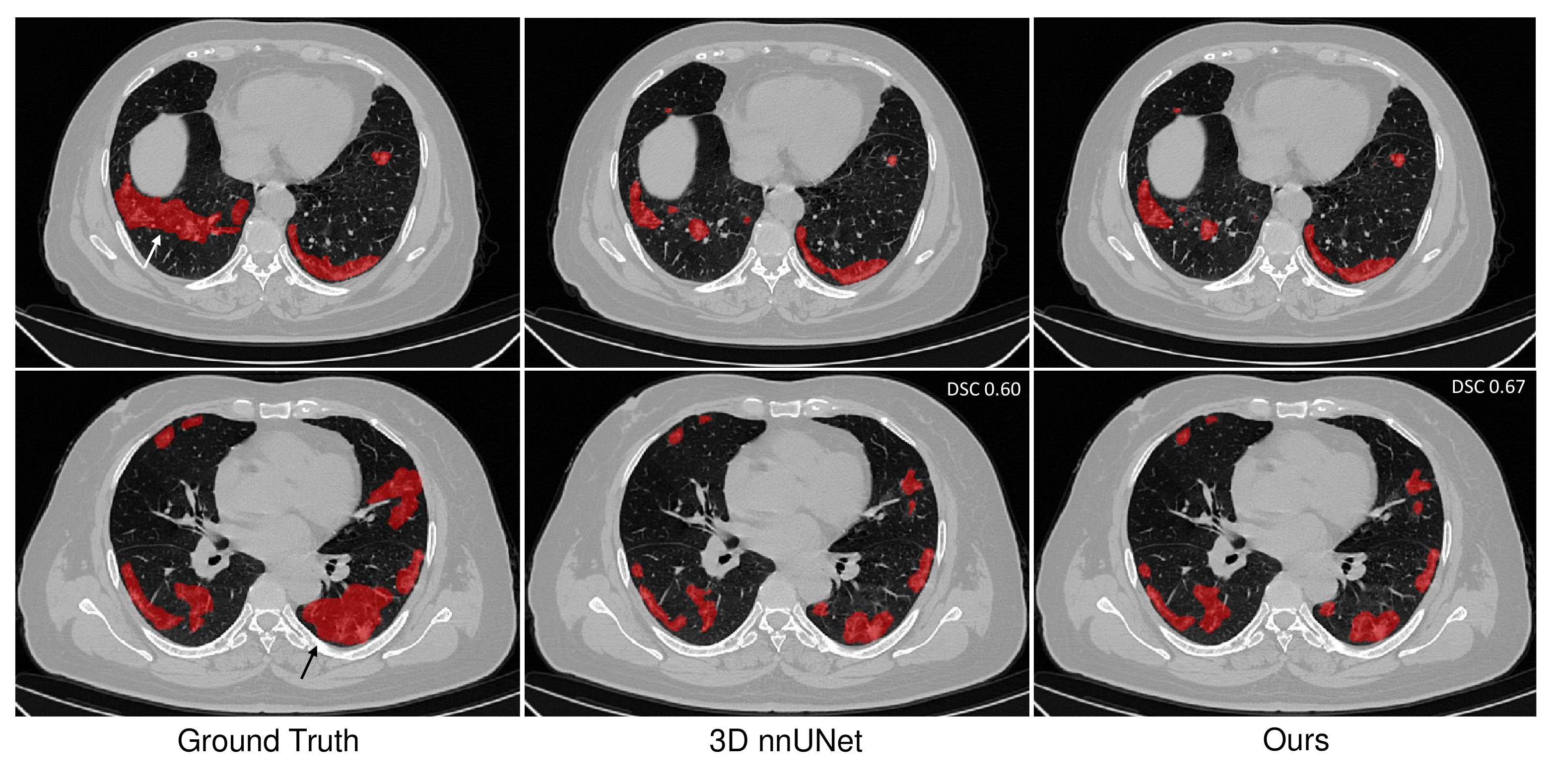}
	\centering
	\caption{Example of challenging cases for COVID-19 lung infection segmentation with limited labeled data. }
	\label{challenge}
\end{figure}

Considering that there are several public non-COVID lung lesion segmentation datasets due to other clinical practice, these datasets may serve as potential profit for generalizing useful information to assist in the related COVID-19 infection segmentation task. Some previous studies also highlight the usage of non-COVID lung lesions \cite{wang2020does,ma2020towards}. However, these existing approaches merely focus on investigating the transferability in COVID-19 infection segmentation. Although their results reveal benefits of pre-training on non-COVID datasets, the improvement is limited when shared knowledge between COVID-19 and non-COVID lung lesions cannot be fully utilized. 
Our experiment reveal that the proposed collaborative learning scheme can effectively exploit shared semantic information by regularizing the consistency between extracted features and promote the training procedure in the absence of sufficient high-quality COVID-19 data.
In addition, our scheme can be extended to utilize unlabeled COVID-19 data for feature relation regularization. 
Experimental results show that even without annotations, our method can use unlabeled scans to explore feature relation and achieve more consistent and robust learning.
Fig.\ref{challenge} presents an example of challenging cases for COVID-19 lung infection segmentation. Although our method can achieve significant improvement by exploiting knowledge from non-COVID lesions, the limitation still exists. We observe that comparing with ground truth, there are still some mis-segmented areas when encountering challenging cases with multiple irregular infections.
As a near future work, we intend to explore how to achieve more robust and reliable knowledge transfer. In addition, we also plan to extend our method to other medical image segmentation tasks to explore the usage of out-of-domain datasets for annotation-efficient deep learning, thus enhancing the applicability of these methods in real-world applications.

\section{Conclusion}

In this paper, we propose a novel relation-driven collaborative learning model to exploit shared knowledge from non-COVID lesions for annotation-efficient COVID-19 CT lung infection segmentation. Specifically, the model consists of two encoders with the same architecture and a shared decoder. 
The general encoder is adopted to capture general lung lesion features based on multiple non-COVID lesions and the target encoder is adopted to focus on task-specific feature of COVID-19 infection. To exploit shared knowledge from non-COVID lesions, we develop a collaborative learning scheme to regularize the relation between extracted features of given input for the training.
We present a set of experiments on 2D slices and 3D volumes based on three COVID-19 datasets and two non-COVID datasets. Experimental results reveal clear benefits of utilizing non-COVID lesions in the absence of sufficient COVID-19 annotations to train a robust segmentation model. 
Moreover, we provide a semi-supervised learning solution to utilize the unlabeled COVID-19 cases for feature relation regularization and achieved performance improvements.
Among all comparison experiments, our proposed method outperforms state-of-the-art methods and illustrates strong potential for real-world applications in the global fight against COVID-19.

%\section*{Acknowledgment}

\bibliographystyle{IEEEtran}
\bibliography{COVIDref}

\begin{thebibliography}{10}
\providecommand{\url}[1]{#1}
\csname url@rmstyle\endcsname
\providecommand{\newblock}{\relax}
\providecommand{\bibinfo}[2]{#2}
\providecommand\BIBentrySTDinterwordspacing{\spaceskip=0pt\relax}
\providecommand\BIBentryALTinterwordstretchfactor{4}
\providecommand\BIBentryALTinterwordspacing{\spaceskip=\fontdimen2\font plus
\BIBentryALTinterwordstretchfactor\fontdimen3\font minus
  \fontdimen4\font\relax}
\providecommand\BIBforeignlanguage[2]{{%
\expandafter\ifx\csname l@#1\endcsname\relax
\typeout{** WARNING: IEEEtran.bst: No hyphenation pattern has been}%
\typeout{** loaded for the language `#1'. Using the pattern for}%
\typeout{** the default language instead.}%
\else
\language=\csname l@#1\endcsname
\fi
#2}}

\bibitem{mazzone2020management}
P.~J. Mazzone, M.~K. Gould, D.~A. Arenberg, A.~C. Chen, H.~K. Choi, F.~C.
  Detterbeck, \emph{et~al.}, ``Management of lung nodules and lung cancer
  screening during the covid-19 pandemic: Chest expert panel report,''
  \emph{Chest}, 2020.

\bibitem{oudkerk2020diagnosis}
M.~Oudkerk, H.~R. B{\"u}ller, D.~Kuijpers, N.~van Es, S.~F. Oudkerk, T.~C.
  McLoud, \emph{et~al.}, ``Diagnosis, prevention, and treatment of
  thromboembolic complications in covid-19: report of the national institute
  for public health of the netherlands,'' \emph{Radiology}, p. 201629, 2020.

\bibitem{li2020coronavirus}
Y.~Li and L.~Xia, ``Coronavirus disease 2019 (covid-19): role of chest ct in
  diagnosis and management,'' \emph{American Journal of Roentgenology}, vol.
  214, no.~6, pp. 1280--1286, 2020.

\bibitem{chung2020ct}
M.~Chung, A.~Bernheim, X.~Mei, N.~Zhang, M.~Huang, X.~Zeng, \emph{et~al.}, ``Ct
  imaging features of 2019 novel coronavirus (2019-ncov),'' \emph{Radiology},
  vol. 295, no.~1, pp. 202--207, 2020.

\bibitem{fang2020sensitivity}
Y.~Fang, H.~Zhang, J.~Xie, M.~Lin, L.~Ying, P.~Pang, \emph{et~al.},
  ``Sensitivity of chest ct for covid-19: comparison to rt-pcr,''
  \emph{Radiology}, p. 200432, 2020.

\bibitem{li2020ct}
K.~Li, Y.~Fang, W.~Li, C.~Pan, P.~Qin, Y.~Zhong, \emph{et~al.}, ``Ct image
  visual quantitative evaluation and clinical classification of coronavirus
  disease (covid-19),'' \emph{European Radiology}, pp. 1--10, 2020.

\bibitem{shi2020review}
F.~Shi, J.~Wang, J.~Shi, Z.~Wu, Q.~Wang, Z.~Tang, \emph{et~al.}, ``Review of
  artificial intelligence techniques in imaging data acquisition, segmentation
  and diagnosis for covid-19,'' \emph{IEEE Reviews in Biomedical Engineering},
  2020.

\bibitem{joskowicz2019inter}
L.~Joskowicz, D.~Cohen, N.~Caplan, and J.~Sosna, ``Inter-observer variability
  of manual contour delineation of structures in ct,'' \emph{European
  Radiology}, vol.~29, no.~3, pp. 1391--1399, 2019.

\bibitem{bilic2019liver}
P.~Bilic, P.~F. Christ, E.~Vorontsov, G.~Chlebus, H.~Chen, Q.~Dou,
  \emph{et~al.}, ``The liver tumor segmentation benchmark (lits),'' \emph{arXiv
  preprint arXiv:1901.04056}, 2019.

\bibitem{heller2019state}
N.~Heller, F.~Isensee, K.~H. Maier-Hein, X.~Hou, C.~Xie, F.~Li, \emph{et~al.},
  ``The state of the art in kidney and kidney tumor segmentation in
  contrast-enhanced ct imaging: Results of the kits19 challenge,''
  \emph{Medical Image Analysis}, vol.~67, p. 101821, 2019.

\bibitem{bernard2018deep}
O.~Bernard, A.~Lalande, C.~Zotti, F.~Cervenansky, X.~Yang, P.-A. Heng,
  \emph{et~al.}, ``Deep learning techniques for automatic mri cardiac
  multi-structures segmentation and diagnosis: is the problem solved?''
  \emph{IEEE Transactions on Medical Imaging}, vol.~37, no.~11, pp. 2514--2525,
  2018.

\bibitem{ma2021abdomenct}
J.~Ma, Y.~Zhang, S.~Gu, C.~Zhu, C.~Ge, Y.~Zhang, \emph{et~al.}, ``Abdomenct-1k:
  Is abdominal organ segmentation a solved problem?'' \emph{IEEE Transactions
  on Pattern Analysis and Machine Intelligence}, 2021.

\bibitem{huang2020serial}
L.~Huang, R.~Han, T.~Ai, P.~Yu, H.~Kang, Q.~Tao, \emph{et~al.}, ``Serial
  quantitative chest ct assessment of covid-19: Deep-learning approach,''
  \emph{Radiology: Cardiothoracic Imaging}, vol.~2, no.~2, p. e200075, 2020.

\bibitem{sun2020adaptive}
L.~Sun, Z.~Mo, F.~Yan, L.~Xia, F.~Shan, Z.~Ding, \emph{et~al.}, ``Adaptive
  feature selection guided deep forest for covid-19 classification with chest
  ct,'' \emph{IEEE Journal of Biomedical and Health Informatics}, vol.~24,
  no.~10, pp. 2798--2805, 2020.

\bibitem{zhang2020clinically}
K.~Zhang, X.~Liu, J.~Shen, Z.~Li, Y.~Sang, X.~Wu, \emph{et~al.}, ``Clinically
  applicable ai system for accurate diagnosis, quantitative measurements, and
  prognosis of covid-19 pneumonia using computed tomography,'' \emph{Cell},
  2020.

\bibitem{wang2020contrastive}
Z.~Wang, Q.~Liu, and Q.~Dou, ``Contrastive cross-site learning with redesigned
  net for covid-19 ct classification,'' \emph{IEEE Journal of Biomedical and
  Health Informatics}, vol.~24, no.~10, pp. 2806--2813, 2020.

\bibitem{ouyang2020dual}
X.~Ouyang, J.~Huo, L.~Xia, F.~Shan, J.~Liu, Z.~Mo, \emph{et~al.},
  ``Dual-sampling attention network for diagnosis of covid-19 from community
  acquired pneumonia,'' \emph{IEEE Transactions on Medical Imaging}, 2020.

\bibitem{wang2020does}
Y.~Wang, Y.~Zhang, Y.~Liu, J.~Tian, C.~Zhong, Z.~Shi, \emph{et~al.}, ``Does
  non-covid19 lung lesion help? investigating transferability in covid-19 ct
  image segmentation,'' \emph{arXiv preprint arXiv:2006.13877}, 2020.

\bibitem{cheplygina2019not}
V.~Cheplygina, M.~de~Bruijne, and J.~P. Pluim, ``Not-so-supervised: a survey of
  semi-supervised, multi-instance, and transfer learning in medical image
  analysis,'' \emph{Medical Image Analysis}, vol.~54, pp. 280--296, 2019.

\bibitem{tajbakhsh2020embracing}
N.~Tajbakhsh, L.~Jeyaseelan, Q.~Li, J.~N. Chiang, Z.~Wu, and X.~Ding,
  ``Embracing imperfect datasets: A review of deep learning solutions for
  medical image segmentation,'' \emph{Medical Image Analysis}, p. 101693, 2020.

\bibitem{van2020survey}
J.~E. Van~Engelen and H.~H. Hoos, ``A survey on semi-supervised learning,''
  \emph{Machine Learning}, vol. 109, no.~2, pp. 373--440, 2020.

\bibitem{ji2019scribble}
Z.~Ji, Y.~Shen, C.~Ma, and M.~Gao, ``Scribble-based hierarchical weakly
  supervised learning for brain tumor segmentation,'' in \emph{International
  Conference on Medical Image Computing and Computer-Assisted
  Intervention}.\hskip 1em plus 0.5em minus 0.4em\relax Springer, 2019, pp.
  175--183.

\bibitem{huang2020multi}
R.~Huang, Y.~Zheng, Z.~Hu, S.~Zhang, and H.~Li, ``Multi-organ segmentation via
  co-training weight-averaged models from few-organ datasets,'' in
  \emph{International Conference on Medical Image Computing and
  Computer-Assisted Intervention}.\hskip 1em plus 0.5em minus 0.4em\relax
  Springer, 2020, pp. 146--155.

\bibitem{zhang2020dodnet}
J.~Zhang, Y.~Xie, Y.~Xia, and C.~Shen, ``Dodnet: Learning to segment
  multi-organ and tumors from multiple partially labeled datasets,''
  \emph{arXiv preprint arXiv:2011.10217}, 2020.

\bibitem{shan2020lung}
F.~Shan, Y.~Gao, J.~Wang, W.~Shi, N.~Shi, M.~Han, \emph{et~al.}, ``Lung
  infection quantification of covid-19 in ct images with deep learning,''
  \emph{arXiv preprint arXiv:2003.04655}, 2020.

\bibitem{amyar2020multi}
A.~Amyar, R.~Modzelewski, H.~Li, and S.~Ruan, ``Multi-task deep learning based
  ct imaging analysis for covid-19 pneumonia: Classification and
  segmentation,'' \emph{Computers in Biology and Medicine}, p. 104037, 2020.

\bibitem{xie2020relational}
W.~Xie, C.~Jacobs, J.-P. Charbonnier, and B.~Van~Ginneken, ``Relational
  modeling for robust and efficient pulmonary lobe segmentation in ct scans,''
  \emph{IEEE transactions on medical imaging}, vol.~39, no.~8, pp. 2664--2675,
  2020.

\bibitem{zhou2020automatic}
T.~Zhou, S.~Canu, and S.~Ruan, ``An automatic covid-19 ct segmentation network
  using spatial and channel attention mechanism,'' \emph{arXiv preprint
  arXiv:2004.06673}, 2020.

\bibitem{zheng2020deep}
C.~Zheng, X.~Deng, Q.~Fu, Q.~Zhou, J.~Feng, H.~Ma, \emph{et~al.}, ``Deep
  learning-based detection for covid-19 from chest ct using weak label,''
  \emph{medRxiv}, 2020.

\bibitem{fan2020inf}
D.-P. Fan, T.~Zhou, G.-P. Ji, Y.~Zhou, G.~Chen, H.~Fu, \emph{et~al.},
  ``Inf-net: Automatic covid-19 lung infection segmentation from ct images,''
  \emph{IEEE Transactions on Medical Imaging}, 2020.

\bibitem{wang2020noise}
G.~Wang, X.~Liu, C.~Li, Z.~Xu, J.~Ruan, H.~Zhu, \emph{et~al.}, ``A noise-robust
  framework for automatic segmentation of covid-19 pneumonia lesions from ct
  images,'' \emph{IEEE Transactions on Medical Imaging}, vol.~39, no.~8, pp.
  2653--2663, 2020.

\bibitem{yao2020label}
Q.~Yao, L.~Xiao, P.~Liu, and S.~K. Zhou, ``Label-free segmentation of covid-19
  lesions in lung ct,'' \emph{arXiv preprint arXiv:2009.06456}, 2020.

\bibitem{ma2020active}
J.~Ma, Z.~Nie, C.~Wang, G.~Dong, Q.~Zhu, J.~He, \emph{et~al.}, ``Active contour
  regularized semi-supervised learning for covid-19 ct infection segmentation
  with limited annotations,'' \emph{Physics in Medicine \& Biology}, vol.~65,
  no.~22, p. 225034, 2020.

\bibitem{chouhan2020novel}
V.~Chouhan, S.~K. Singh, A.~Khamparia, D.~Gupta, P.~Tiwari, C.~Moreira,
  \emph{et~al.}, ``A novel transfer learning based approach for pneumonia
  detection in chest x-ray images,'' \emph{Applied Sciences}, vol.~10, no.~2,
  p. 559, 2020.

\bibitem{majeed2020covid}
T.~Majeed, R.~Rashid, D.~Ali, and A.~Asaad, ``Covid-19 detection using cnn
  transfer learning from x-ray images,'' \emph{medRxiv}, 2020.

\bibitem{misra2020multi}
S.~Misra, S.~Jeon, S.~Lee, R.~Managuli, I.-S. Jang, and C.~Kim, ``Multi-channel
  transfer learning of chest x-ray images for screening of covid-19,''
  \emph{Electronics}, vol.~9, no.~9, p. 1388, 2020.

\bibitem{ronneberger2015u}
O.~Ronneberger, P.~Fischer, and T.~Brox, ``U-net: Convolutional networks for
  biomedical image segmentation,'' in \emph{International Conference on Medical
  Image Computing and Computer-Assisted Intervention}.\hskip 1em plus 0.5em
  minus 0.4em\relax Springer, 2015, pp. 234--241.

\bibitem{cciccek20163d}
{\"O}.~{\c{C}}i{\c{c}}ek, A.~Abdulkadir, S.~S. Lienkamp, T.~Brox, and
  O.~Ronneberger, ``3d u-net: learning dense volumetric segmentation from
  sparse annotation,'' in \emph{International Conference on Medical Image
  Computing and Computer-Assisted Intervention}.\hskip 1em plus 0.5em minus
  0.4em\relax Springer, 2016, pp. 424--432.

\bibitem{liu2020semi}
Q.~Liu, L.~Yu, L.~Luo, Q.~Dou, and P.~A. Heng, ``Semi-supervised medical image
  classification with relation-driven self-ensembling model,'' \emph{IEEE
  Transactions on Medical Imaging}, 2020.

\bibitem{gatys2016neural}
L.~Gatys, A.~Ecker, and M.~Bethge, ``A neural algorithm of artistic style,''
  \emph{Journal of Vision}, vol.~16, no.~12, pp. 326--326, 2016.

\bibitem{dou20163d}
Q.~Dou, H.~Chen, Y.~Jin, L.~Yu, J.~Qin, and P.-A. Heng, ``3d deeply supervised
  network for automatic liver segmentation from ct volumes,'' in
  \emph{International Conference on Medical Image Computing and
  Computer-Assisted Intervention}.\hskip 1em plus 0.5em minus 0.4em\relax
  Springer, 2016, pp. 149--157.

\bibitem{ma2020towards}
J.~Ma, Y.~Wang, X.~An, C.~Ge, Z.~Yu, J.~Chen, \emph{et~al.}, ``Towards
  data-efficient learning: A benchmark for covid-19 ct lung and infection
  segmentation,'' \emph{Medical physics}, 2020.

\bibitem{simpson2019large}
A.~L. Simpson, M.~Antonelli, S.~Bakas, M.~Bilello, K.~Farahani,
  B.~Van~Ginneken, \emph{et~al.}, ``A large annotated medical image dataset for
  the development and evaluation of segmentation algorithms,'' \emph{arXiv
  preprint arXiv:1902.09063}, 2019.

\bibitem{paszke2019pytorch}
A.~Paszke, S.~Gross, F.~Massa, A.~Lerer, J.~Bradbury, G.~Chanan, \emph{et~al.},
  ``Pytorch: An imperative style, high-performance deep learning library,'' in
  \emph{Advances in Neural Information Processing Systems}, 2019, pp.
  8026--8037.

\bibitem{isensee2020nnu}
F.~Isensee, P.~F. Jaeger, S.~A. Kohl, J.~Petersen, and K.~H. Maier-Hein,
  ``nnu-net: a self-configuring method for deep learning-based biomedical image
  segmentation,'' \emph{Nature Methods}, pp. 1--9, 2020.

\bibitem{laine2016temporal}
S.~Laine and T.~Aila, ``Temporal ensembling for semi-supervised learning,''
  \emph{arXiv preprint arXiv:1610.02242}, 2016.

\bibitem{oktay2018attention}
O.~Oktay, J.~Schlemper, L.~L. Folgoc, M.~Lee, M.~Heinrich, K.~Misawa,
  \emph{et~al.}, ``Attention u-net: Learning where to look for the pancreas,''
  \emph{arXiv preprint arXiv:1804.03999}, 2018.

\bibitem{schlemper2019attention}
J.~Schlemper, O.~Oktay, M.~Schaap, M.~Heinrich, B.~Kainz, B.~Glocker,
  \emph{et~al.}, ``Attention gated networks: Learning to leverage salient
  regions in medical images,'' \emph{Medical image analysis}, vol.~53, pp.
  197--207, 2019.

\bibitem{li2018h}
X.~Li, H.~Chen, X.~Qi, Q.~Dou, C.-W. Fu, and P.-A. Heng, ``H-denseunet: hybrid
  densely connected unet for liver and tumor segmentation from ct volumes,''
  \emph{IEEE transactions on medical imaging}, vol.~37, no.~12, pp. 2663--2674,
  2018.

\bibitem{zhou2019unet++}
Z.~Zhou, M.~M.~R. Siddiquee, N.~Tajbakhsh, and J.~Liang, ``Unet++: Redesigning
  skip connections to exploit multiscale features in image segmentation,''
  \emph{IEEE transactions on medical imaging}, vol.~39, no.~6, pp. 1856--1867,
  2019.

\bibitem{roth2021rapid}
H.~Roth, Z.~Xu, C.~T. Diez, R.~S. Jacob, J.~Zember, J.~Molto, \emph{et~al.},
  ``Rapid artificial intelligence solutions in a pandemic-the covid-19-20 lung
  ct lesion segmentation challenge,'' 2021.

\end{thebibliography}

\end{document}